\documentclass[12pt]{iopart}
\usepackage{iopams,epsfig}

\def\fl{}

\def\Secref#1{Section~\ref{#1}}

\def\secref#1{section~\ref{#1}}

\def\ssecref#1{section~\ref{#1}}
\def\ssecsref#1#2{sections~\ref{#1} and~\ref{#2}}
\def\apref#1{\ref{#1}}

\def\figref#1{figure~\ref{#1}}

\def\const{{\rm const}}

\def\bea{\begin{eqnarray}}
\def\eea{\end{eqnarray}}

\def\phi{\varphi}

\def\({\left(}
\def\){\right)}
\def\<{\left\langle}
\def\>{\right\rangle}
\def\l{\left}
\def\r{\right}
\def\la{\langle}
\def\ra{\rangle}

\def\d{\partial}
\def\Dt{{\rmd\over\rmd t}\,}
\def\diff{\rmd}

\def\vx{\bi{x}}
\def\vy{\bi{y}}
\def\vk{\bi{k}}
\def\vu{\bi{u}}
\def\vB{\bi{B}}

\def\vf{\bi{f}}

\def\ud{u_\nu}
\def\Wd{W_\nu}
\def\Weq{W_0}
\def\teq{t_0}
\def\tres{t_\eta}

\def\krms{k_{\rm rms}}

\def\ks{k_{\rm s}} 
\def\CK{C_{\rm K}}
\def\vA{v_{\rm A}}

\def\kpar{k_{\parallel}}

\def\lpar{\ell_{\parallel}}
\def\lperp{\ell_{\perp}}

\def\ls{\ell_{\rm s}}

\def\tcorr{\tau_{\rm c}}
\def\teddy{\tau_{\rm eddy}}
\def\gKA{{\bar\gamma}}
\def\imnu{{\tilde\nu}}
\def\cg{c_1}
\def\cW{c_2}

\def\kIR{k_*}
\def\xiIR{\xi_*}

\def\kd{k_{\nu}}

\def\kf{k_f}
\def\kres{k_{\eta}}

\def\kreskin{k_{\eta}^{\rm (kin)}}

\def\Re{{Re}}

\def\Pr{{Pr}}

\def\Bsq{{\langle B^2 \rangle}}

\begin{document}

\footnotesize

\title[A model of the nonlinear turbulent dynamo]{A model of 
nonlinear evolution and saturation of the turbulent MHD dynamo}

\author{A~A~Schekochihin, S~C~Cowley, 
G~W~Hammett\footnote[1]{Permanent address: 
Princeton Plasma Physics Laboratory, P.~O.~Box~451, 
Princeton, NJ~08543, USA}, 
J~L~Maron\footnote[2]{Present address: 
Department of Physics and Astronomy, 
University of Rochester, Rochester, NY~14627, USA}}
\address{Plasma Physics Group, Imperial College, 
Blackett Laboratory, Prince Consort Road, London~SW7~2BW, England}
\author{J~C~McWilliams}
\address{Department of Atmospheric Sciences, 
UCLA, Los Angeles, CA 90095-1565, USA}
\eads{\mailto{a.schekochihin@ic.ac.uk}, 
\mailto{steve.cowley@ic.ac.uk}, 
\mailto{hammett@princeton.edu},
\mailto{maron@tapir.caltech.edu}, 
\mailto{jcm@atmos.ucla.edu}}

\begin{abstract}
The growth and saturation of magnetic field in 
conducting turbulent media with large magnetic Prandtl numbers 
are investigated. This regime is very common in 
low-density hot astrophysical plasmas. During the early (kinematic) stage, 
weak magnetic fluctuations grow exponentially and concentrate 
at the resistive scale, which lies far below the hydrodynamic 
viscous scale. The evolution becomes nonlinear when the magnetic 
energy is comparable to the kinetic energy of the viscous-scale 
eddies. A physical picture of the ensuing nonlinear evolution 
of the MHD dynamo is proposed. Phenomenological considerations 
are supplemented with a simple Fokker--Planck model 
of the nonlinear evolution of the magnetic-energy spectrum. 
It is found that, while the shift of the bulk of the magnetic energy 
from the subviscous scales to the velocity scales may be possible, 
it occurs very slowly --- at the resistive, 
rather than dynamical, time scale (for galaxies, 
this means that generation of large-scale magnetic fields 
cannot be explained by this mechanism). 
The role of Alfv\'enic motions and the implications for 
the fully developed isotropic MHD turbulence are discussed. 

\vskip1cm 

\noindent
24 July 2002\\\\
\fbox{{\em New Journal of Physics}~{\bf 4}~84~(2002) ({\tt http://www.njp.org})}\\\\
E-print {\tt astro-ph/0207503}

\end{abstract}

\maketitle

\section{Introduction}
\label{sec_intro}

It has long been appreciated~\cite{Batchelor_dynamo} that 
a generic three-dimensional turbulent flow of a conducting 
fluid is likely to be a {\em dynamo,} i.e., it amplifies  
an initial weak seed magnetic field 
provided the magnetic Reynolds number is above a certain 
instability threshold (usually between~10 and~100). 
The amplification is exponentially fast and occurs 
due to stretching of the 
magnetic field lines by the random velocity shear 
associated with the turbulent eddies. 
The same mechanism leads to an equally rapid 
decrease of the field's spatial coherence scale, as the 
stretching and folding by the ambient flow brings the field 
lines with oppositely directed fields 
ever closer together~\cite{Ott_review,SCMM_folding,SMCM_structure}. 

If the magnetic Prandtl number ($\Pr=\nu/\eta$, the ratio 
of the viscosity and the magnetic diffusivity of the fluid) 
is large and the hydrodynamic turbulence has Kolmogorov form, 
the scale of the magnetic 
field can decrease by a factor of~$\Pr^{1/2}$ below the 
viscous scale of the fluid. This situation is known as 
{\em the Batchelor regime}~\cite{Batchelor_regime} and is characterized 
by the magnetic-energy concentration at subviscous scales. 
This MHD regime is very common in astrophysical plasmas, which
tend to have very low density and high temperature. 
Examples include warm interstellar medium, 
some accretion discs, 
jets,
protogalaxies, 
intracluster gas in galaxy clusters,
early Universe,~etc. 
(a representative set of recent references is~\cite{KA,Kulsrud_review,SBK_review,Balbus_Hawley_review,Heinz_Begelman,Kulsrud_etal_proto,Malyshkin_clusters,Narayan_Medvedev,Son,Christensson_Hindmarsh_Brandenburg}).
The object of our principal interest are magnetic fields of galaxies, 
often thought to have been generated by the turbulent dynamo in the 
interstellar medium. The challenge is to reconcile 
the preponderance of small-scale magnetic fluctuations 
resulting from the weak-field (kinematic) stage of the dynamo 
with the observed galactic fields coherent 
at scales of approximately 1~kpc, or 10~times larger than the 
outer scale of the interstellar turbulence, which is forced 
by supernova explosions at scales 
of~$\sim100$~pc (\cite{Beck_review,Widrow_review,Han_Wielebinski} 
are some of the recent reviews of relevant observations). 

The galactic magnetic fields generally have energies comparable 
to the energy of the turbulent motions of the interstellar medium, 
and, therefore, cannot be considered weak. Indeed, the 
growth of the magnetic energy established for the kinematic regime 
inevitably leads to the field becoming strong enough to resist 
further amplification and to modify the properties of 
the ambient turbulence by exerting a force (Lorentz tension) 
on the fluid. This marks the transition from the kinematic (linear) 
stage of the dynamo to the nonlinear regime. 
In the astrophysical context, the main issue is what happens to 
the coherence scale of the field during the nonlinear stage 
and, specifically, whether a mechanism could be envisioned 
that would effectively transfer the magnetic energy from 
small (subviscous) to velocity scales: perhaps as large 
as the outer scale of the turbulence and above. 

A large amount of work has been devoted to this issue during 
the last 50~years (a small subset of the relevant references 
is~\cite{Batchelor_dynamo,Pouquet_Frisch_Leorat,Meneguzzi_Frisch_Pouquet,Kida_Yanase_Mizushima,Chandran_closure,Cho_Vishniac,Brandenburg,Chou,MCM_dynamo}). 
Since most phenomenological theories of the {\em steady-state} 
MHD turbulence envision a state of eventual detailed 
scale-by-scale equipartition between magnetic and velocity energies~\cite{Iroshnikov,Kraichnan_IK,Goldreich_Sridhar_strong}, 
it has been widely expected that the dynamo 
would converge to such a state via some form 
of inverse cascade of magnetic energy. The implication would be 
an eventual dominant magnetic-energy concentration at the outer scale 
of the turbulence --- a convenient starting point 
for the operation of the helical mean-field 
dynamo~\cite{Moffatt,Blackman_review} 
or of the helical inverse cascade~\cite{Frisch_etal,Pouquet_Frisch_Leorat}.
Note that we believe the inverse cascade of magnetic helicity that takes place 
in helical MHD turbulence~\cite{Frisch_etal,Pouquet_Frisch_Leorat} 
to be unrelated to the small-scale processes we discuss here: 
we are interested in ways to shift the magnetic energy from subviscous 
scales to the outer (forcing) scale of the turbulence, 
while the helical inverse cascade transfers magnetic helicity 
from the forcing scale to even larger scales. 
In the galactic dynamo theory, helicity-related 
effects should not be relevant for the small-scale turbulence because 
they operate at the time scale associated with the overall rotation 
of the galaxy, which is 10 times larger than the turnover time 
of the largest turbulent eddies ($10^8$ and $10^7$~years, 
respectively~\cite{KA}). 

To prevent confusion, let us spell out that by {\em turbulent dynamo} 
we mean the turbulent amplification of the energy of magnetic 
{\em fluctuations}, not of the mean field. Fluctuation-dynamo 
theories describe magnetic fields at the scales of the turbulence, 
mean-field-dynamo theories refer to much larger scales 
(the mean field is usually the field averaged over the flow scales). 
While fast mean-field dynamo 
is not possible in an isotropic system without net helicity 
(see, e.g.,~\cite{Moffatt}), the fluctuation dynamo 
is~\cite{Batchelor_dynamo,Kazantsev,KA}. Moreover, fluctuation 
dynamos in three-dimensional turbulent flows are usually fast, i.e., 
proceed at a finite rate in the limit of arbitrarily 
small magnetic diffusivity. 

Thus,
in this work, we study the possibility of the nonhelical, 
nonlinear, large-$\Pr$ 
fluctuation dynamo saturating with magnetic energy concentrated 
at the outer scale of the turbulence. Using a simple spectral 
Fokker--Planck model inspired by the quantitative theory 
of the kinematic stage of the dynamo and by a physical model 
of the nonlinear stage, 
we establish a possible mechanism for shifting the bulk 
of the magnetic energy from the subviscous scales to the 
velocity scales. The energy transfer mechanism proposed here 
involves selective resistive decay of small-scale fields 
combined with continued amplification of the field at larger 
scales.
Therefore, the time needed for this process to complete itself 
turns out to be the resistive time associated with the velocity 
scales of the turbulence, {\it not} a dynamical time. 
In large-$\Pr$ systems, this time is enormously long and, 
in fact, can easily exceed the age of the Universe. 
Thus, a steady state with magnetic energy at velocity scales, 
even if attainable in principle, is not relevant for such systems. 
However, the scenario of the evolution and saturation 
of the nonlinear dynamo proposed here has fundamental implications 
for our understanding of the structure of the forced isotropic 
MHD turbulence and of the results of the recent numerical 
experiments. 

The plan of the rest of this paper is as follows. 
In \secref{sec_kinematic}, we review the kinematic theory. 
It is worked out in some detail because it constitutes the mathematical 
framework of the subsequent nonlinear extension 
presented in \secref{sec_FP}. 
In \secref{sec_phenom}, a phenomenological theory of the nonlinear 
dynamo is proposed. This section contains 
our main physical argument and is, therefore, the part of 
this paper that a reader not interested in details should peruse 
before jumping to the discussion section at the end. 
In \secref{sec_FP} we proceed to formulate a Fokker--Planck 
model of the evolution of the magnetic-energy spectrum. 
The model is then solved numerically and analytically. Self-similar 
solutions are obtained. 
Both the qualitative physics of \secref{sec_phenom} and 
the analytical solutions of \secref{sec_FP} 
exhibit a resistively controlled effective 
transfer of the magnetic energy to the velocity scales. 
The resulting long-time asymptotic state of the fully developed isotropic 
MHD turbulence is discussed in \ssecref{ssec_MHD}. 
\Secref{sec_conclusions} contains a discussion of our findings and of 
the issues left open in our approach. 

\section{The kinematic dynamo}
\label{sec_kinematic}

\subsection{The growth of magnetic energy}

The kinematic dynamo is described by the induction equation 
\bea
\label{ind_eq}
\d_t\vB + \vu\cdot\bnabla\vB = \vB\cdot\bnabla\vu + \eta\Delta\vB,
\eea
where $\vB$~is the magnetic field, $\eta$~is the magnetic diffusivity, 
and $\vu$~is the random incompressible velocity field, whose statistics are 
externally prescribed. The following {\em exact} 
evolution law for the magnetic energy~$W={1\over2}\Bsq$ is an immediate 
corollary of~\eref{ind_eq}: 
\bea
\label{W_eq_exact}
\Dt W = 2\gamma(t) W - 2\eta\krms^2(t) W,
\eea
where, by definition, 
\bea
\fl \gamma(t)={\la\vB\vB:\bnabla\vu\ra \over \Bsq},\qquad 
\krms^2(t) = {\la\(\d_j B^i\)\(\d_j B^i\)\ra \over \Bsq}
= {1\over W}\int_0^\infty\rmd k\,k^2 M(t,k), 
\eea 
and $M(t,k)={1\over2}\int\rmd\Omega_\vk\la|\vB(t,\vk)|^2\ra$ is 
the $k$-shell-integrated magnetic-energy spectrum. 
The angle brackets signify ensemble averages, but, the system being 
homogeneous, can also be thought of as volume integrals. 
Batchelor~\cite{Batchelor_dynamo} argued that, 
in a turbulent flow, $\gamma(t)$~would tend to 
some constant~$\gKA$, thus implying exponential growth 
of the magnetic energy. This is essentially 
a consequence of the line-stretching 
property of turbulence~\cite{Cocke,Orszag}. 

The growth of the magnetic energy is accompanied 
by the decrease of the characteristic scale of the magnetic fluctuations, 
which also proceeds exponentially fast in time at 
a rate~$\sim\gKA$~\cite{Batchelor_dynamo}. 
In a large-$\Pr$ medium, this process quickly 
transfers the bulk of the magnetic energy to scales below 
the hydrodynamic viscous cutoff. 
In the Kolmogorov turbulence, the fastest eddies 
are the viscous-scale ones, so it is their turnover time 
that determines the time scale at which the small-scale kinematic 
dynamo operates: $\gKA\sim\kd\ud$, where $\kd$ and $\ud$~are the 
characteristic wave number and velocity of these eddies~\cite{KA}. 

\subsection{The magnetic-energy spectrum}
\label{ssec_spectrum}

Since the subviscous-scale magnetic fields ``see'' the ambient 
flow as a random linear shear, one might conjecture that their 
statistics should not be very sensitive to the specific 
structure of the flow, provided the latter is sufficiently random 
in space and time. This gives a physical 
rationale for modelling the ambient turbulence with some 
synthetic random velocity field whose statistics would be 
simple enough to render the problem analytically solvable. 
The most eminently successful such scheme has been 
the Kazantsev--Kraichnan model of passive 
advection~\cite{Kazantsev,Kraichnan_ensemble} which uses 
a Gaussian white-in-time velocity as a stand-in for the 
real turbulence: 
\bea
\label{KK_field}
\<u^i(t,\vx)u^j(t',\vx')\> = \delta(t-t')\kappa^{ij}(\vx-\vx'). 
\eea
On the basis of this model, it is possible to show that 
the magnetic-energy spectrum at subviscous 
scales ($k\gg\kd$) satisfies the following equation~\cite{Kazantsev,Vainshtein_SSF-1,Vainshtein_SSF-2,KA,Gruzinov_Cowley_Sudan,SBK_review} 
\bea
\label{SSF_eq}
\d_t M = {\gKA\over5}\(k^2{\d^2 M\over\d k^2} 
- 2k{\d M\over\d k} + 6 M\) - 2\eta k^2 M, 
\eea
where~$\gKA=-(1/6)\l[\Delta\kappa^{ii}(\vy)\r]_{\vy=0}$,
which is consistent with its physical meaning as the characteristic 
eddy-turnover rate of the turbulence~\cite{SK_tcorr,SBK_review}. 

Note that, in order for equation~\eref{SSF_eq} to be an adequate 
description of the spectral properties of the magnetic field, 
it must be a conservative approximation, i.e., be consistent 
with the exact energy-evolution law~\eref{W_eq_exact}. 
Indeed, we can rewrite equation~\eref{SSF_eq} in 
a conservative Fokker--Planck form 
\bea
\label{FP_form}
\d_t M = {\d\over\d k}\l[D(k){\d M\over\d k} - V(k) M\r] + 2\gKA M 
- 2\eta k^2 M, 
\eea
where $D(k)=\gKA k^2/5$ is the diffusion coefficient 
and $V(k)=4\gKA k/5$ the ``drift velocity'' 
in $k$ space ($V>0$ implies the spectrum moving towards higher~$k$). 
We see that $W(t)=\int_0^\infty\rmd k\,M(t,k)$ satisfies~\eref{W_eq_exact} 
with~$\gamma(t)=\gKA$.\\ 

\noindent
{\em Remarks on the validity of the Fokker--Planck model.} 
Equation \eref{SSF_eq} is valid in the limit that the flow
velocity has a zero correlation time as given by~\eref{KK_field}, 
which implies an infinite kinetic-energy
density. In order to calculate the relevant parameters of this model for
comparison with the more realistic situation where the flow
has a finite correlation time, one may take the following steps. 
In~\eref{KK_field}, $\delta(t-t')$ is replaced by 
$(1/2\tcorr) \exp(-|t-t'|/\tcorr)$, where 
$\tcorr$ is the correlation time of the velocity. 
Doing a calculation parallel to that in~\cite{SK_tcorr}, one gets 
\bea
\label{gamma_d}
\gKA = {2\tcorr\over3}\int_0^\infty\rmd k\,k^2 E(k),
\eea 
where $E(k)$ is the energy spectrum of the fluid turbulence. 
[For a more refined model, one might allow for the decorrelation 
rate~$\tcorr^{-1}$ to vary with $k$ 
and remain inside the integral in~\eref{gamma_d}, 
or use a three-mode decorrelation rate instead to 
account for a finite ratio of the velocity and magnetic-field 
correlation times. But as long as one uses a value for
$\tcorr$ characteristic of the fastest (highest-$k$) eddies, which dominate
the integral in~\eref{gamma_d}, and $\tcorr$ does not become large compared
to the magnetic-field correlation time, the main scaling is the same.] 
Note that the integral in~\eref{gamma_d} is proportional to the square
of the characteristic eddy-turnover rate~$\teddy^{-1}$ of the flow. 
It is clear that, in a fully developed turbulence, $\tcorr\sim\teddy$, 
so we may rewrite \eref{gamma_d} as 
\bea
\label{gamma_d2}
\gKA = \cg\l[\int_0^\infty\rmd k\,k^2 E(k)\r]^{1/2},
\eea 
where $\cg$ is a coefficient of order unity.  

The other main assumption in~\eref{SSF_eq} is that we are
focusing on magnetic excitations at~$k\gg\kd$, 
which raises the question of the appropriate boundary
conditions or modifications to the theory at low~$k$. The effect the
large-scale structure of the velocity field may have on the statistics 
of the small-scale magnetic field is incompletely 
understood. However, when the bulk of the magnetic energy is at 
the subviscous scales, it is probably safe to assume that the spectrum
is not significantly affected by the details of what happens 
at large scales (see discussion in~\cite{SBK_review}). 
For~$\Pr\gg1$, 
we shall find that~$\krms\gg\kd$, except perhaps at very large~$t$. 
In this regime, our results are insensitive to the precise form 
of the boundary condition at small~$k$, and we can simply 
specify a suitably low boundary $\kIR\ll\krms$ across which there 
is no flux of magnetic energy. 

At~$k\lesssim\kd$, there is no scale separation between 
the velocity and magnetic field, the local (in $k$~space) 
approximation~\eref{SSF_eq} breaks down, and 
one must, strictly speaking, use a more general 
integrodifferential equation for~$M(t,k)$~\cite{Kazantsev,KA,SBK_review}, 
but simplifications may be possible. 
In the Kolmogorov picture of isotropic turbulence, the
dominant stretching of eddies at a particular scale is due to
interactions with other eddies at comparable scales. This suggests that
the effect of small-scale velocity fields on larger-scale magnetic 
fields can be neglected, i.e., that the ``turbulent diffusivity'' 
acting on the magnetic field at some $k<\kd$ is dominated 
by the contribution from the velocities at the same~$k$. 
As an approximation, one would then just 
continue to use the Fokker--Planck equation~\eref{FP_form} but with the
expression~\eref{gamma_d2} for~$\gKA$ modified to 
account for the fact that velocities at a given scale are not very 
effective in causing net stretching of magnetic fields at larger~scales: 
\bea
\label{gamma_k}
\gKA(k) = \cg\l[\int_0^k\rmd k'k^{\prime2} E(k')\r]^{1/2}.
\eea 
For $k<\kd$, the dominant contribution to this integral
comes from $k'\sim k$, i.e., from eddies of sizes comparable to 
the scale of the field that is being stretched. 
Note that $\gKA(k)$ in~\eref{gamma_k} has the same scaling 
as the $k$-dependent eddy-damping rate in the popular phenomenological 
EDQNM closure~\cite{Orszag_EDQNM}, 
where setting $\cg = 0.36$ results in a good fit to the Kolmogorov
constant~\cite{Pouquet_Frisch_Leorat}. 
The shearing rate $\gKA(k)$ drops to zero for $k$ below 
the forcing wavenumber~$\kf$ of the turbulence, so the ``infrared''
(IR) cutoff~$\kIR$ 
just needs to be low enough that $\gKA(\kIR)$ [and thus the flux of
magnetic energy in $k$~space, the expression in square brackets 
in equation~\eref{FP_form}] also vanishes. 
In the analytic solutions presented below, we shall neglect any 
$k$~dependence of $\gKA$ for simplicity 
and write the results in terms of a general IR cutoff~$\kIR$. 
One could set $\kIR\sim\kf$ as a crude model of the argument above, 
but the precise value of~$\kIR$ will not affect any of our results 
until \ssecref{ssec_steady_state}, where 
the choice and the meaning of~$\kIR$ will be revisited. 

\subsection{The diffusion-free (ideal-MHD) regime} 
\label{ssec_diff_free}

Let us discuss how a magnetic excitation initially concentrated 
at some wave numbers~$k'$ such that~$\kd\ll k'\ll\kres$ will evolve 
with time. Neglecting the resistive-diffusion term for the 
time being ($\eta=0$), 
we immediately find the Green's-function solution of~\eref{SSF_eq}: 
\bea
\label{M_lognorm}
\fl M(t,k) = \rme^{(3/4)\gKA t}\int_0^\infty{\diff k'\over k'}\,M_0(k') 
\({k\over k'}\)^{3/2} {1\over\sqrt{(4/5)\pi\gKA t}}\,
\exp\(-{\bigl[\ln(k/k')\bigr]^2\over(4/5)\gKA t}\),
\eea
where $M_0(k')$~is the initial spectrum. 
This solution shows that 
(i) the amplitude of each Fourier mode grows exponentially in time 
at the rate~$(3/4)\gKA$, 
(ii) the number of excited modes 
[i.e., the width of the lognormal envelope in~\eref{M_lognorm}] 
also grows exponentially at the rate $(4/5)\gKA$, 
(iii) the peak of the excitation originating from each initially 
present mode~$k'$ moves towards larger~$k$: 
$k_{\rm peak} = k'\exp\l[(3/5)\gKA t\r]$, leaving 
a power spectrum~$\sim k^{3/2}$ behind. 
The cumulative effect is 
the total magnetic-energy growth at the rate~$2\gKA$. 

The solution~\eref{M_lognorm} becomes invalid 
when the magnetic excitation reaches 
the velocity scale~$\kd^{-1}$ and/or the resistive scale~$\kres^{-1}$. 
In the former case, including the $k$~dependence of~$\gKA$  
as discussed at the end of \ssecref{ssec_spectrum} 
would be the minimal adjustment necessary to model large-scale effects. 
However, the bulk of the magnetic energy is at subviscous scales and 
is not significantly affected by such modifications.  
We shall, therefore, continue to rely on~\eref{SSF_eq} 
and ask what happens when the resistive scale is reached. 

\subsection{The resistive regime}
\label{ssec_resistive}

The asymptotic form of the magnetic-energy spectrum 
in the resistive regime can be determined by solving 
an eigenvalue problem.
Seeking the solution in the form~\cite{KA} 
\bea
\label{M_ev}
M(t,k) = \rme^{\lambda\gKA t} \Phi(\xi),\qquad \xi=k/\kres,\qquad
\kres=(\gKA/10\eta)^{1/2}, 
\eea
we~get 
\bea
\label{Bessel_eq}
\xi^2\Phi'' - 2\xi\Phi' + (6-5\lambda)\Phi - \xi^2\Phi = 0.
\eea
The solutions of~\eref{Bessel_eq} are Bessel functions. 
Demanding that $\Phi(\xi\to\infty)=0$, we~find 
\bea
\label{Phi_sln}
\Phi(\xi) = \const\,\xi^{3/2} K_{\nu(\lambda)}(\xi),\qquad 
\nu(\lambda)=\sqrt{5\(\lambda-3/4\)}, 
\eea
where $K_{\nu}$~is the Macdonald function. 

The eigenvalue~$\lambda$ 
must be determined from the boundary condition at small~$k$. 
This can be done in several different 
ways leading to the same result in the asymptotic case~$\Pr\gg1$
(see, e.g.,~\cite{Gruzinov_Cowley_Sudan,SBK_review} 
and the exhaustive reference list in~\cite{SBK_review}). 
Here we choose some finite IR cutoff~$\kIR$ 
and impose a zero-flux boundary condition at~$k=\kIR$.
This gives [see equation~\eref{FP_form}]
\bea
\label{zero_flux}
\xiIR\Phi'(\xiIR)-4\Phi(\xiIR) = 0,\qquad \xiIR=\kIR/\kres.
\eea
Substituting~\eref{Phi_sln} yields the following transcendental 
equation for~$\lambda=\lambda(\xiIR)$: 
\bea
\label{lambda_eq}
\fl \xiIR K'_{\nu(\lambda)}(\xiIR) - {5\over2}\, K_{\nu(\lambda)}(\xiIR) 
= \(\nu(\lambda)-{5\over2}\r)K_{\nu(\lambda)}(\xiIR) 
- \xiIR K_{\nu(\lambda)+1}(\xiIR) = 0.
\eea
It is immediately clear that \eref{lambda_eq} has no solutions 
with real~$\nu(\lambda)$, i.e., with~$\lambda\ge3/4$. Indeed, 
such solutions could only exist if the first term were positive: 
$\nu(\lambda)>5/2$, which would imply~$\lambda>2$. Since the total 
energy cannot grow faster than at the rate~$2\gKA$, 
$\lambda>2$~is impossible. Now, the Macdonald 
function~$K_{\nu(\lambda)}(\xi)$ 
of imaginary order~$\nu(\lambda)=\rmi\imnu(\lambda)$ 
oscillates as it approaches~$\xi=0$. 
Since the spectrum must be a nonnegative function, 
all zeros of~$K_{\rmi\imnu(\lambda)}(\xi)$ must lie to the left of 
the IR cutoff~$\xiIR$: 
\bea
\label{condn_zeros}
z_{\rmi\imnu(\lambda),m}\le\xiIR.
\eea
As we are considering a large-$\Pr$ case, we must 
take~$\xiIR=\kIR/\kres\sim\Pr^{-1/2}\ll 1$. 
The condition~\eref{condn_zeros} then implies~$\imnu(\lambda)\ll 1$. 
In these limits, \eref{lambda_eq} reduces to 
\bea
\label{zero_M}
\fl K_{\rmi\imnu(\lambda)}(\xiIR)\simeq 
-{1\over\imnu(\lambda)}\,\sin\l[\imnu(\lambda)\ln(\xiIR/2)\r] = 0,\qquad 
\imnu(\lambda)=\sqrt{5(3/4-\lambda)}.
\eea
Thus, in the large-$\Pr$ limit, the zero-flux boundary 
condition~\eref{zero_flux} is equivalent to requiring that the spectrum 
should vanish at the IR~cutoff: $\Phi(\xiIR)=0$ (cf.~\cite{SBK_review}).
Taking the first (rightmost) zero in~\eref{zero_M}, 
so that \eref{condn_zeros}~is observed, we find the desired eigenvalue
\bea
\label{lambda_sln}
\lambda \simeq {3\over4} - {\pi^2\over5\l[\ln(\xiIR/2)\r]^2} 
= {3\over4} - \Or\({1\over\ln^2(\Pr^{1/2})}\).
\eea

For~$\xi\gg\xiIR$, the spectrum is well described by~\eref{Phi_sln}
with~$\nu(\lambda)=0$.
Thus, in the kinematic regime with diffusion, the magnetic-energy 
spectrum has a constant profile with a $k^{3/2}$ scaling in the 
range~$\kd\ll k\ll\kres$ [see~\eref{Phi_sln}], 
and all modes grow exponentially 
at the same rate~$\simeq (3/4)\gKA$~\cite{KA}. Most importantly, 
the bulk of the magnetic energy is concentrated at the resistive 
scales~$\sim\kres^{-1}$. 

\section{A phenomenological theory of the nonlinear dynamo}
\label{sec_phenom}

\subsection{The onset of nonlinearity} 
\label{ssec_onset}

When the energy of the magnetic field reaches values such that 
the back reaction on the flow cannot be neglected, 
the dynamo becomes nonlinear and the 
velocity~$\vu$ can no longer be considered as given. It has to be 
determined self-consistently from the Navier--Stokes equation: 
\bea
\label{NS_eq}
\d_t\vu + \vu\cdot\bnabla\vu = \nu\Delta\vu 
-\bnabla p + \vB\cdot\bnabla\vB + \vf,
\eea
where $\nu$~is the fluid viscosity, 
$\vf$~is some random large-scale forcing, 
$p$~is the pressure determined from the incompressibility 
condition~$\bnabla\cdot\vu=0$, and $p$ and $\vB$ have been normalized 
by $\rho$ and $(4\pi\rho)^{1/2}$, respectively, $\rho$~being 
the (constant) density of the medium. 


From equation~\eref{NS_eq}, it is evident that the nonlinear 
regime is characterized by at least partial balancing of 
the inertial and magnetic-tension terms: 
$\vu\cdot\bnabla\vu\sim\vB\cdot\bnabla\vB$. 
In order to correctly estimate the tension force~$\vB\cdot\bnabla\vB$, 
it is {\em not} enough to know the energy of the field 
and its characteristic scale as determined from the isotropic 
spectra in \secref{sec_kinematic}. Indeed, the gradient 
in~$\vB\cdot\bnabla\vB$ is locally aligned with the field direction, 
so one must invoke the geometrical structure of the field
lines. As was shown in our earlier 
studies~\cite{SCMM_folding,SMCM_structure,MCM_dynamo}, the small-scale 
fields generated by the kinematic dynamo are organized in 
folds: the smallness of their characteristic scale is due 
to rapid spatial-direction reversals, while the field lines 
remain mostly straight up the scale of the stretching eddy. 
It follows that, during the kinematic stage, 
$\vB\cdot\bnabla\vB\sim\kd B^2$. 
Since the inertial term is~$\vu\cdot\bnabla\vu\sim\kd\ud^2$, 
the nonlinearity becomes important when 
the small-scale magnetic energy equalizes with the energy of 
the viscous-scale eddies: 
$B^2\sim\ud^2$, or $W\sim\Wd$~\cite{SCMM_folding}. 
Note that if the folded structure of the field had been ignored,
then estimating $\vB\cdot\bnabla\vB\sim\kres B^2$ 
would have led to the incorrect conclusion that nonlinearities 
become important at the much lower magnetic energies: 
$B^2\sim\ud^2/\Pr^{1/2}$.

\subsection{The nonlinear-growth stage} 
\label{ssec_nlin_growth}

The magnetic back reaction must then
act to suppress the random shear flows associated with 
these eddies. However, the next-larger-scale eddies still 
have energies larger than that of the magnetic field, 
though the turnover rate of these eddies is smaller. We conjecture 
that these eddies will continue to amplify the field in the same, 
essentially kinematic, fashion. 
The scale of these eddies is~$\ls>\kd^{-1}$, the folds 
are elongated accordingly (see remark in \secref{ssec_first}), 
and the tension force is~$\vB\cdot\bnabla\vB\sim B^2/\ls$. 
The corresponding inertial term is~$\vu\cdot\bnabla\vu\sim u_{\ls}^2/\ls$, 
so these eddies become suppressed when~$B^2\sim u_{\ls}^2$, 
whereupon it will be the turn of the next-larger eddies to provide 
the dominant stretching action. At all times, it is the smallest 
unsuppressed eddy that most rapidly stretches the field. 
Thus, we define the stretching scale~$\ls(t)$ at a given 
time~$t$~by 
\bea
\label{nlin_balance}
W(t)\sim {1\over2}\,u^2_{\ls(t)}. 
\eea
In equation~\eref{W_eq_exact}, $\gamma(t)$~is now the turnover 
rate of the eddies of scale~$\ls(t)$: 
$\gamma(t)\sim u_{\ls(t)}/\ls(t)$. Then 
$\gamma(t) W(t) \sim u^3_{\ls(t)}/\ls(t) \sim \epsilon$, 
where $\epsilon=\const$~is the Kolmogorov energy flux. 
Equation~\eref{W_eq_exact} becomes 
\bea
\label{W_eq_eps}
\Dt W \simeq \chi\epsilon - 2\eta\krms^2(t) W, 
\eea
where $\chi$~is some constant of order unity. 
The physical meaning of this equation is as follows. 
The turbulent energy injected at the forcing scale cascades 
hydrodynamically down to the scale~$\ls(t)$ where 
a finite fraction~$\chi$ of it is 
diverted into the small-scale magnetic fields. 
Equation~\eref{W_eq_eps} implies that the magnetic energy 
should grow linearly with time during this stage: $W(t)\sim\epsilon t$. 
Furthermore, comparing~\eref{W_eq_eps} and~\eref{W_eq_exact}, 
we get 
\bea
\label{gamma_eps}
\gamma(t)\simeq{\chi\epsilon\over 2W(t)}\simeq{\alpha\over t}, 
\eea 
where $\alpha$~is a constant order-one coefficient to be determined 
in~\ssecref{ssec_first}. 

We stress that the field is still organized in folds of 
characteristic parallel length~$\ls(t)$ with direction 
reversals perpendicular to~$\vB$ at the resistive scale. Note that 
the resistive scale 
now changes with time: a straightforward estimate 
gives $\kres(t)\sim\l[\gamma(t)/\eta\r]^{1/2}\sim(\eta t)^{-1/2}$. 
Thus, the resistive cutoff moves to larger scales. 
Indeed, as the stretching slows down with decreasing~$\gamma(t)$, 
{\em selective decay} eliminates the modes at the high-$k$ end 
of the spectrum for which the resistive time is now shorter 
than the stretching time and which, consequently, cannot 
be sustained anymore. In \ssecref{ssec_first}, 
we shall see that the magnetic-energy spectrum 
evolves in a self-similar fashion during this stage. 

\begin{figure}[t]
\centerline{\epsfig{file=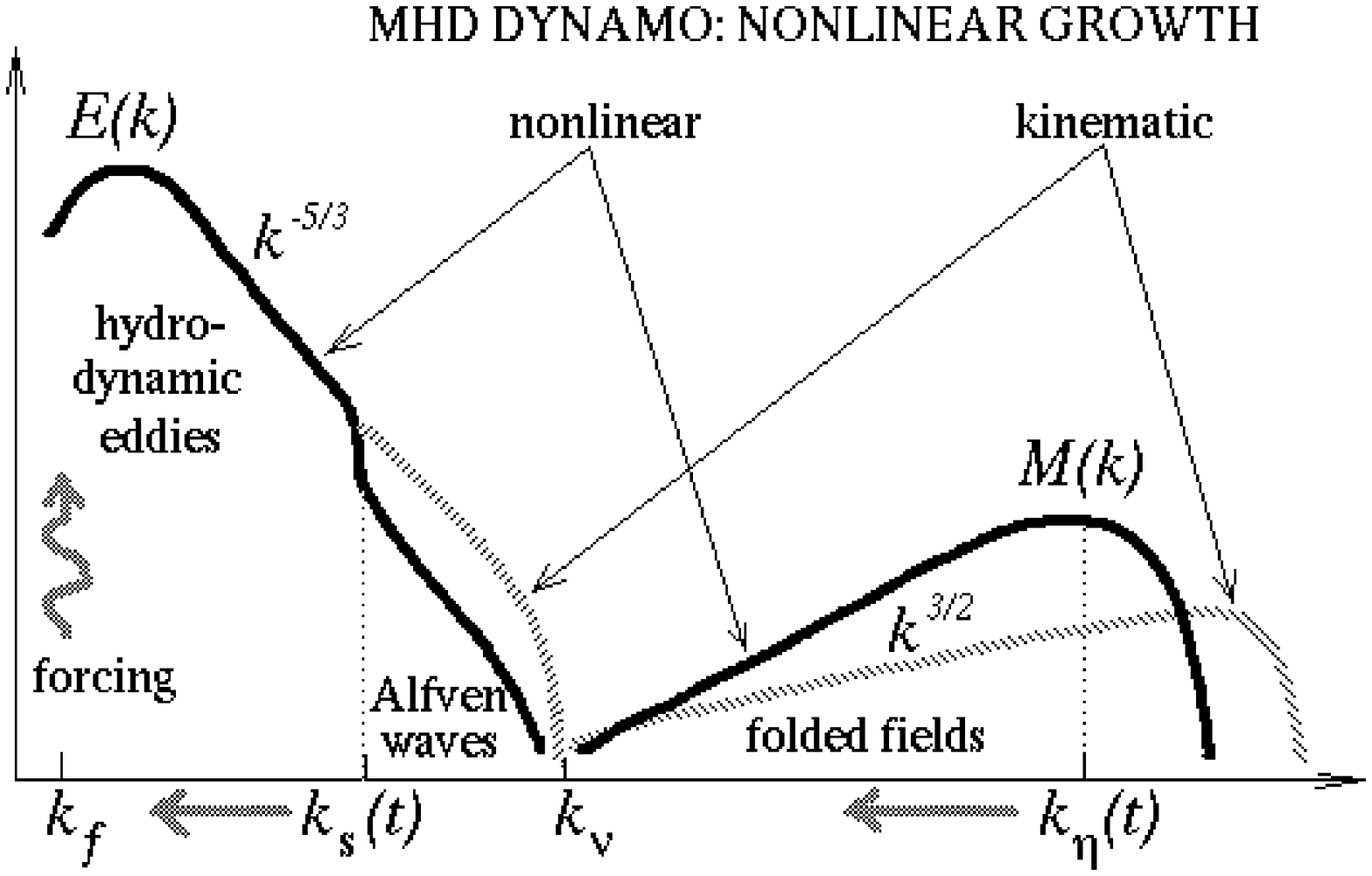,width=3.25in}
\epsfig{file=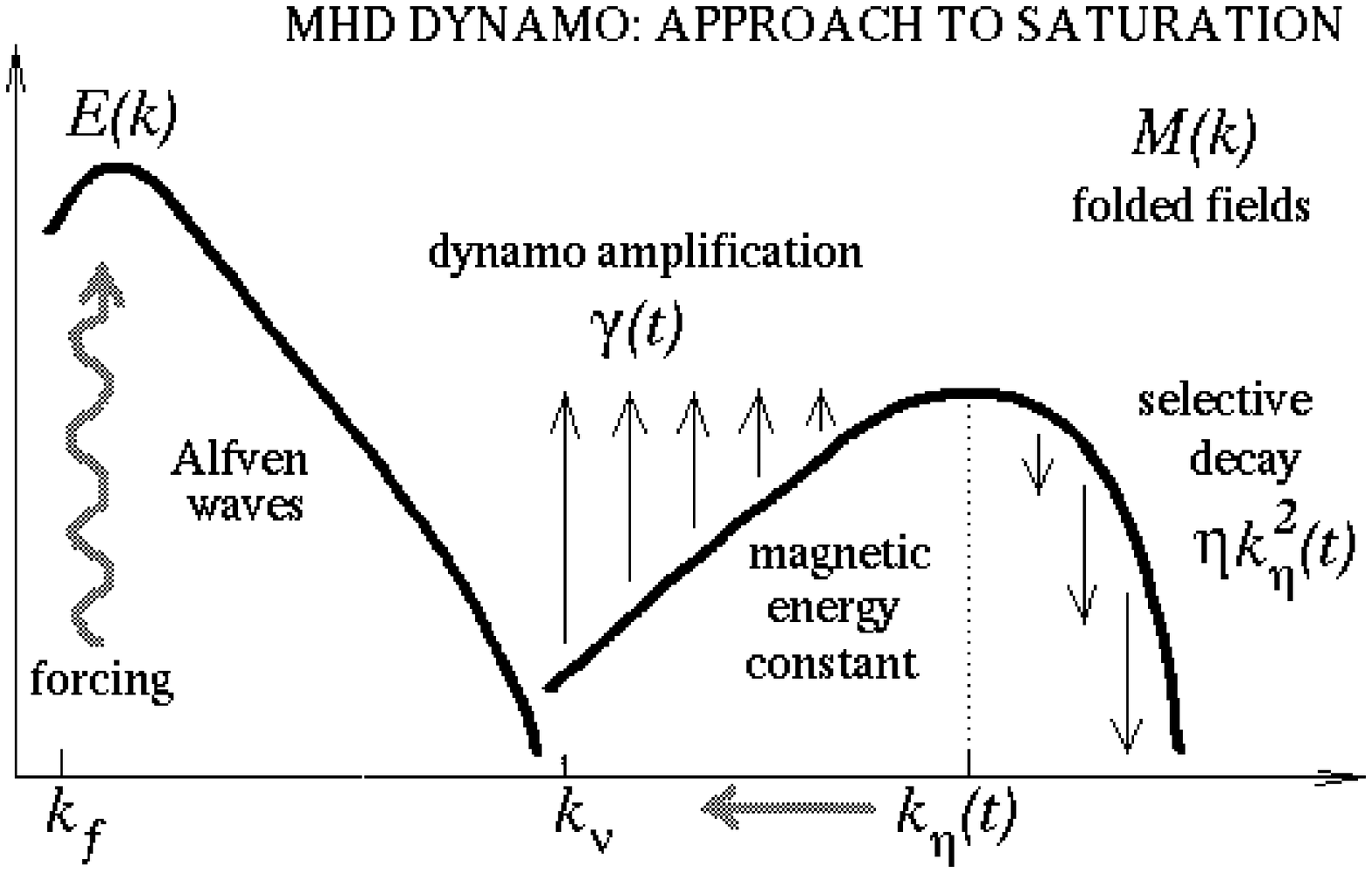,width=3.25in}}
\caption{\label{fig_spectra} Sketches of the 
kinetic and magnetic energy spectra in the nonlinear regime. 
{\em Left:} the nonlinear-growth 
stage (\ssecref{ssec_nlin_growth}). 
During the kinematic stage, the velocity spectrum $E(k)$ is Kolmogorov, 
the magnetic spectrum $M(k)$ is peaked 
at~$\kres\sim\Pr^{1/2}\kd$. In the nonlinear-growth stage, 
the magnetic field is amplified by the eddies at 
the stretching scale~$\ls\sim\ks^{-1}$. They are the eddies 
whose energy is equal to the total magnetic energy 
at any given time~$t$. 
The motions with~$k<\ks(t)$ are still hydrodynamic, 
but those with~$k>\ks(t)$ are Alfv\'enic. 
As the magnetic energy grows, both $\ks$ and $\kres$ decrease. 
{\em Right:} the approach to saturation (\ssecref{ssec_approach}). 
Selective decay eliminates high-$k$ modes, $\kres$~decreases, 
the stretching rate $\gamma\sim\eta\kres^2$ decreases 
accordingly, so the magnetic energy stays constant.} 
\end{figure}

Note that some fluid motions do survive at scales 
below~$\ls(t)$ and down to the viscous cutoff. These are the 
motions that do not amplify the magnetic field. 
They are waves of Alfv\'enic nature propagating along the 
folds of direction-reversing magnetic fields 
(\apref{ap_waves}). A finite fraction of the hydrodynamic 
energy arriving from the large scales is channelled into 
the turbulence of these waves (\figref{fig_spectra}). 
Since the Alfv\'enic motions are dissipated {\em viscously}, not 
resistively, we shall assume that they do not have 
a secular effect on the folded small-scale field, i.e., 
that they do not significantly change its scale or energy 
(see \apref{ap_waves} and further discussion in \ssecref{ssec_MHD}). 

The process of scale-by-scale suppression of the stretching motions 
continues until the magnetic energy 
becomes comparable to the energy~$\Weq$ of the outer-scale eddies (i.e., 
to the energy of the turbulence) and $\ls\sim\kf^{-1}$, the forcing 
scale. The time scale for this to happen 
is easily estimated to be the turnover time of the outer-scale 
eddies~$\teq\sim(\kf u_0)^{-1}\sim\gKA^{-1}\Re^{1/2}$, 
where $u_0$~is the velocity of these eddies. 
The resistive cutoff scale is now larger than during 
the kinematic stage: 
$\kres(\teq)/\kreskin\sim (\teq\gKA)^{-1/2}\sim\Re^{-1/4}$,  
where by $\kreskin$ we denote the resistive cutoff in the kinematic 
regime. However, this is not a very significant shift: 
for example, in the interstellar medium, where~$\Re\sim10^4$, 
it is by a meager factor of~$0.1$. Recall that $\kd/\kf\sim\Re^{3/4}$ 
and $\kreskin/\kd\sim\Pr^{1/2}$, 
so $\kres(\teq)/\kf\sim(\Re\,\Pr)^{1/2}$,
i.e., the scale separation between the 
field-generating motions and the resulting magnetic fields is 
actually increased. The magnetic field is still 
small scale and arranged in folds of length~$\kf^{-1}$ with 
direction reversals at the resistive scale. 

\subsection{The approach to saturation}
\label{ssec_approach}

At this point, the hydrodynamic energy stirred up at the forcing scale 
is transfered directly into the small-scale folded magnetic field 
(i.e., in~\eref{W_eq_exact}, $\gamma\sim\kf u_0$). 
As there are no scales in the system larger than $\kf^{-1}$, 
there can be no further growth of the magnetic energy. 
Indeed, the tension in the folds is now 
$\vB\cdot\bnabla\vB\sim\kf W$ and 
the force balance in~\eref{NS_eq} requires 
$\vB\cdot\bnabla\vB\sim\vu\cdot\bnabla\vu\sim\kf\Weq$, so $W\sim \Weq$. 
Note that setting~$\rmd W/\rmd t=0$ in~\eref{W_eq_exact} 
gives $\gamma(t)=\eta\krms^2(t)$.  
This is a reflection of the fact that the resistive cutoff~$\kres\sim\krms$ 
is now determined by the zero-sum 
balance between the field amplification (stretching) 
and the resistive dissipation: 
\bea
\label{res_balance}
\gamma(t)\sim\eta\kres^2(t). 
\eea 
However, at scales larger than the resistive cutoff, 
i.e., for $k<\kres$, the field-amplification rate~$\gamma$ is larger 
than the resistive-dissipation rate~$\eta k^2$ 
(\figref{fig_spectra}, right panel). 
Since the total 
magnetic energy is not allowed to grow anymore, the growth of 
modes with $k<\kres$ must be compensated by the decrease 
in the value of~$\kres$, i.e., by further selective decay 
of the high-$k$ modes. This, in turn, should lead to 
further decrease of~$\gamma$ in accordance with the 
resistive balance~\eref{res_balance}.
We shall see in \ssecref{ssec_second} that the evolution 
of the magnetic-energy spectrum in this regime is likely 
to be self-similar with $\gamma(t)\sim1/t$ 
and $\kres(t)\sim(\eta t)^{-1/2}$. 

Thus, we expect that this last 
stage of the nonlinear dynamo should be characterized by $\gamma$ 
dropping below the turnover rate~$\kf u_0$ of the outer-scale eddies. 
Therefore, the rate of the energy transfer into the small-scale magnetic 
field decreases, as this channel of energy dissipation 
becomes inefficient. 
Instead, we expect the energy injected by the forcing to be 
increasingly diverted into the Alfv\'enic turbulence 
that is left throughout the inertial range in the wake of the 
suppression of the stretching motions. 

\subsection{The fully developed isotropic MHD turbulence}
\label{ssec_MHD}

We saw in the preceding sections that 
$\kres$ decreases both in the nonlinear-growth 
stage and during the subsequent slower approach to saturation. 
It is very important to understand how far the decrease 
of~$\kres$ can proceed.  
We recall that, in our arguments so far, we have disregarded 
the Alfv\'enic component of the turbulence. 
It is not, however, justified to do so 
{\em after} the small-scale magnetic energy reaches the velocity scales.
Indeed, the decrease of~$\kres(t)$ is basically a consequence 
of the balance between the field-amplification and 
the resistive-dissipation terms in the energy equation~\eref{W_eq_exact}: 
as $\gamma(t)$ drops, so does~$\kres(t)\sim\krms(t)$; 
once the characteristic scale of the small-scale 
fields reaches the viscous scale~$\kd^{-1}$, 
the Alfv\'enic turbulence will start to affect~$\krms$ 
in an essential way: since the waves are damped 
at the viscous scale, $\krms$~must stabilize at $\krms\sim\kd$.  
The resulting turbulent state features folded 
magnetic fields reversing directions at the viscous scale plus 
Alfv\'en waves in the inertial range propagating along the folds. 
There are then two possibilities: either (i)~this represents the 
final steady state of the isotropic MHD turbulence, 
or (ii)~further evolution will lead to unwinding of the 
folds and continued energy transfer to larger scales, 
so that the spectrum will eventually peak at the outer scale and 
have an Alfv\'enic power tail extending through the inertial 
range. The turbulence in the inertial range would 
in the latter case be of the usual Alfv\'en-wave 
kind~\cite{Iroshnikov,Kraichnan_IK,Goldreich_Sridhar_strong}, where 
the large-scale magnetic fluctuations provide a mean field along which 
the inertial-range Alfv\'en waves propagate. 
A generic MHD turbulent steady state would thus 
emerge at the end of the nonlinear-dynamo evolution. 
This dichotomy cannot be resolved at the level 
of the present model and requires further study. 
However, in either case, $\krms\sim\kd$, 
so the statistical steady state of the fully developed isotropic 
MHD turbulence is characterized by the equalization of the 
resistive and viscous scales. 

The time scale at which such an equalization is brought about 
is the resistive time associated with the viscous 
scale of the turbulence:~$\tres(\kd)\sim\(\eta\kd^2\)^{-1}$. 
We immediately notice that, in order for the slow approach 
to saturation described in \ssecref{ssec_approach} to be realizable, 
this time scale must be longer than the turnover time of 
the outer-scale eddies: $\tres(\kd)\gg\teq$, which 
requires~$\Pr\gg\Re^{1/2}$. In other words, the resistive scale 
at the end of the nonlinear-growth stage must still be smaller 
than the viscous scale: $\kres(\teq)/\kd\sim\Pr^{1/2}\Re^{-1/4}\gg1$. 
This constitutes the ``true large-$\Pr$ regime,'' which is 
the one relevant for astrophysical plasmas (thus, in the 
interstellar medium, $\Re\sim10^4$, $\Pr\sim10^{14}$). 
In this regime, the magnetic energy saturates at the equipartition 
level, $W\sim\Weq$. From the energy balance~\eref{W_eq_exact}, 
we get an estimate 
of the amount of turbulent power that is still dissipated 
resistively: $\chi\epsilon\sim \gamma W \sim \eta\kd^2\Weq 
\sim \epsilon\,\Re^{1/2}\Pr^{-1}$. Thus, $\chi\ll1$, i.e., most 
of the injected power now goes into the Alfv\'enic motions 
(which are dissipated viscously). 

The other possibility is $\Pr\lesssim\Re^{1/2}$. 
In this case, the nonlinear-growth stage described 
in \ssecref{ssec_nlin_growth} is curtailed 
with magnetic energy at a subequipartition value: 
\bea
\label{W_subeq}
W\sim {\tres(\kd)\over\teq}\,\Weq \sim {\Pr\over\sqrt{\Re}}\,\Weq,
\eea 
and resistivity continuing to take a significant ($\chi$~is order one) 
part in the dissipation of the turbulent energy. 
If no further evolution takes place [possibility (i) above], 
\eref{W_subeq} gives an estimate of the saturation energy of 
the magnetic component of the turbulence. 
This regime is characterized by Prandtl numbers which 
may still be much larger than unity, but are not large enough 
to capture all of the physics of large-$\Pr$ dynamo. 
Most of the extant numerical simulations appear to 
be in this regime (see \secref{sec_conclusions}).\\

We should now like to examine the implications of the proposed 
scenario in a somewhat more quantitative fashion, 
assuming from now on~$\Pr\gg\Re^{1/2}\gg1$.
A busy reader with no time for these 
slow-paced developments may skip to \secref{sec_conclusions}. 

\section{The Fokker--Planck model of the dynamo}
\label{sec_FP}

\subsection{Formulation of the model} 
\label{ssec_model}

The expression~\eref{gamma_eps} suggests a way to build up the 
phenomenological picture just laid out into a semiquantitative 
model of the evolution of the magnetic spectrum throughout the 
nonlinear regime. 
Let us keep the general form of equation~\eref{SSF_eq}, 
but replace~$\gKA$ by a time-dependent function~$\gamma(t)$, 
for which we stipulate the following {\em ad hoc} expression: 
\bea
\label{gamma_Ek}
\gamma(t) = \cg\l[\int_0^{\ks(t)}\diff k\,k^2 E(k)\r]^{1/2},\qquad
\cW\int_{\ks(t)}^\infty\diff k\, E(k) = W(t),
\eea
where $\cg$ and $\cW$~are some numerical constants of order unity, 
$E(k)$~is the hydrodynamic energy spectrum in the absence 
of magnetic field, $\ks(t)$~is {\em defined} by the second 
expression in~\eref{gamma_Ek}, 
and $W(t)$ is the total magnetic energy at time~$t$.
The Fokker--Planck model we propose for the nonlinear 
evolution of the magnetic spectrum is then
\bea
\label{SSF_eq_nlin}
\d_t M = {\gamma(t)\over5}\(k^2{\d^2 M\over\d k^2} 
- 2k{\d M\over\d k} + 6 M\) - 2\eta k^2 M.  
\eea

We now assume the following crude model profile of the 
hydrodynamic energy spectrum of the Kolmogorov turbulence 
in the absence of magnetic field:
\bea
\label{Ek_model}
E(k) =\cases{
\CK \epsilon^{2/3}k^{-5/3} & for $k\in[\kf,\kd]$,\\
0  & elsewhere,\\}
\eea 
where $\CK\simeq1.5$ is the Kolmogorov constant. 
The value of~$\kd$ is set to enforce 
consistency of the model profile~\eref{Ek_model} with the 
exact relation $\epsilon=2\nu\int_0^\infty\diff k\,k^2 E(k)$: 
\bea
\kd=\kf\(1+{\epsilon^{1/3}\over(3/2)\CK\nu\kf^{4/3}}\)^{3/4} 
\sim \Re^{3/4}\kf.
\eea
Using~\eref{Ek_model}, it is straightforward to~derive from~\eref{gamma_Ek}
\bea
\label{gamma_model}
\fl \gamma(t) = \gKA\l[1 - {1\over\(1+\Weq/\Wd\)^2}\r]^{-1/2} 
\l[{1\over\(1+W(t)/\Wd\)^2} 
- {1\over\(1+\Weq/\Wd\)^2}\r]^{1/2},
\eea
where $\Wd={3\over2}\,\cW\CK\epsilon^{2/3}\kd^{-2/3}$ 
and $\Weq=\cW\int_0^\infty\diff k\,E(k)$  
are of the order of the energies of the viscous- 
and of the outer-scale eddies, respectively. 
Here, $\gKA = c_1(\epsilon/2\nu)^{1/2}$ follows from evaluating
\eref{gamma_Ek} for $W\to0$, $\ks = k_\nu$. 
Note that
\bea 
\label{Weq_over_Wd}
{\Weq\over\Wd} = \({\kd\over\kf}\)^{2/3} - 1 \sim\Re^{1/2}\gg1. 
\eea
The kinematic regime~$\gamma(t)\simeq\gKA$ (\secref{sec_kinematic}), 
the nonlinear-growth regime~\eref{gamma_eps} (\ssecref{ssec_nlin_growth}), 
and the long-time asymptotic~$\gamma(t)\to0$ (\ssecref{ssec_approach}) 
are all contained in the 
model expression~\eref{gamma_model}. 

In the above formulae, $E(k)$~is the hydrodynamic 
turbulence spectrum {\it in the absence of the magnetic field}, 
and we have effectively assumed that, at each time~$t$ during the 
evolution of the dynamo, $E(k)$~remains unaffected 
by the magnetic fields at~$k<\ks(t)$. 
As was explained in \ssecref{ssec_nlin_growth}, 
in the suppressed part of 
the inertial range~$k>\ks(t)$, Alfv\'enic 
turbulence is excited, which draws a finite fraction 
of the turbulent power and, indeed, in the long run, supplants 
the small-scale magnetic fields as the main 
energy-dissipation channel (see \ssecref{ssec_MHD}). 
This part of the picture is 
not included in our Fokker--Planck model~\eref{SSF_eq_nlin}. We believe 
that it can be considered independently of the evolution 
of the small-scale magnetic-energy spectrum as long 
as the bulk of the magnetic energy stays below the viscous scale 
(see \ssecsref{ssec_MHD}{ssec_steady_state}). 


\subsection{Self-similar solutions} 
\label{ssec_ssim}

The Fokker--Planck model~\eref{SSF_eq_nlin} 
leads to the emergence of two distinct 
intermediate regimes where the magnetic-energy spectrum 
evolves in a self-similar fashion. 
In both regimes, the bulk of 
the magnetic energy moves from small to larger scales. 
We stress that the asymptotic 
self-similar solutions presented below are not sensitive to the 
particular form~\eref{gamma_model} of~$\gamma(t)$. 
The essential features 
are those discussed in \ssecsref{ssec_nlin_growth}{ssec_approach}. 
The expression~\eref{gamma_model} will only be used to validate 
our solutions numerically. 

Let us now study equation \eref{SSF_eq_nlin}.
We notice that, if $\gamma(t)$~can be represented as
\bea
\label{gamma_ssim}
\gamma(t)={\alpha\over t},
\eea 
where $\alpha$~is a numerical constant, 
equation~\eref{SSF_eq_nlin} admits a self-similar solution 
\bea
\label{M_ssim}
M(t,k) = {W(t)\over\kres(t)}\,\Phi(\xi),\qquad
\xi=k/\kres(t),\qquad \int_0^\infty\rmd\xi\,\Phi(\xi)=1,\\
\kres(t)=\l[\gamma(t)/10\eta\r]^{1/2} = 
\(\alpha/10\eta t\)^{1/2},\\
\label{W_mu}
W(t) = C t^{\mu},
\eea
where the function $\Phi(\xi)$ satisfies
\bea
\label{Phi_eq_mu}
\xi^2\Phi'' - \(2+{5\over2}\,\alpha^{-1}\)\xi\Phi' 
+ \(6-5\,{2\mu+1\over2}\,\alpha^{-1}\)\Phi - \xi^2\Phi = 0.
\eea
Clearly, for~$\gamma(t)\propto1/t$, 
the appearance of self-similarity is inevitable on dimensional 
grounds because there is no fixed time scale in the problem in this case. 
The parameters $C$ and~$\mu$ in~\eref{W_mu} 
must be determined from the assumptions about the physics of the problem. 
The value of~$\alpha$ is fixed by the solvability of 
equation~\eref{Phi_eq_mu}.

\subsection{The first self-similar regime: nonlinear growth}
\label{ssec_first}

In \ssecref{ssec_nlin_growth}, we argued that $W(t)$ should grow linearly 
with time during the intermediate stage of 
the nonlinear dynamo when the magnetic back reaction gradually 
eliminates the stretching component of the inertial-range eddies. 
Thus, \eref{W_mu} should hold with~$\mu=1$. Then, 
by virtue of~\eref{gamma_eps}, $\gamma(t)$~has the form~\eref{gamma_ssim} 
required for the self-similar solution to exist. 
The model expression~\eref{gamma_model} correctly reproduces 
this regime: substituting~\eref{gamma_ssim} 
into~\eref{gamma_model}, we~get  
\bea
\label{W_first}
W(t) \simeq \alpha^{-1}\gKA\Wd t \simeq {\chi\epsilon\over2\alpha}\,t,\qquad 
\gKA^{-1} \ll t \ll \gKA^{-1}{\Weq\over\Wd} 
\sim\gKA^{-1}\Re^{1/2}\sim\teq.
\eea
Comparing the above asymptotic with equation~\eref{W_eq_eps}, 
we note that $1/2\alpha$~can be interpreted as the fraction of 
the turbulent power received by the magnetic fields 
that actually contributes to the growth of their energy. 
The rest is dissipated resistively. 
The upper time limit~$\teq$ for this regime is the turnover time 
of the outer-scale eddies. 

In order to determine the 
value of~$\alpha$, we have to solve the equation~\eref{Phi_eq_mu} 
with~$\mu=1$ and enforce the right boundary conditions. 
This is an eigenvalue problem very similar to that which we 
considered in~\ssecref{ssec_resistive}. The solution is again 
expressible in terms of the Macdonald function:
\bea
\label{Phi_sln_ssim}
\Phi(\xi) = \const\,\xi^{\sigma(\alpha)} K_{\nu(\alpha)}(\xi),\qquad
\sigma(\alpha) = {3\over2} + {5\over4}\,\alpha^{-1},\\ 
\label{nu_alpha}
\nu(\alpha) = {5\over4}\({1\over\alpha}-{1\over\alpha_+}\)^{1/2}
\({1\over\alpha}-{1\over\alpha_-}\)^{1/2},\qquad 
{1\over\alpha_\pm} = {2\over5}\(\pm4\sqrt{6}-9\).
\eea

It is now necessary to implement the boundary condition at small~$k$. 
Again, we should like to pick some finite IR cutoff~$\kIR$ and 
to impose the zero-flux condition~\eref{zero_flux} 
(see discussion in \secref{ssec_steady_state} on the choice of~$\kIR$). 
However, strictly speaking, introducing the new 
dimensional parameter~$\kIR$ 
breaks the self-similarity of the problem: instead of~\eref{M_ssim}, 
we must now write
\bea
\label{M_nonssim}
M(t,k) = {W(t)\over\kres(t)}\,\Phi(\xi,\xiIR),\qquad
\int_{\xiIR}^\infty\rmd\xi\,\Phi(\xi,\xiIR)=1, 
\eea
where $\xiIR=\kIR/\kres(t)=\kIR(10\eta\alpha^{-1}t)^{1/2}$ is a new
{\em time-dependent} dimensionless variable. 
Since the non-self-similar solution~\eref{M_nonssim} has 
a self-similar asymptotic when $\xiIR\to0$, we can still 
use~\eref{Phi_sln_ssim}, provided~$\xiIR\ll1$. 
This solution remains valid as long as~$t\ll\alpha\tres$, where 
$\tres=(10\eta\kIR^2)^{-1}$ is the resistive time associated with 
the IR~cutoff~$\kIR$. In large-$\Pr$ media, this is usually longer than 
the nonlinear-growth regime itself lasts according to~\eref{W_first} 
(see \ssecsref{ssec_MHD}{ssec_steady_state} though). 

\begin{figure}[t]
\centerline{\epsfig{file=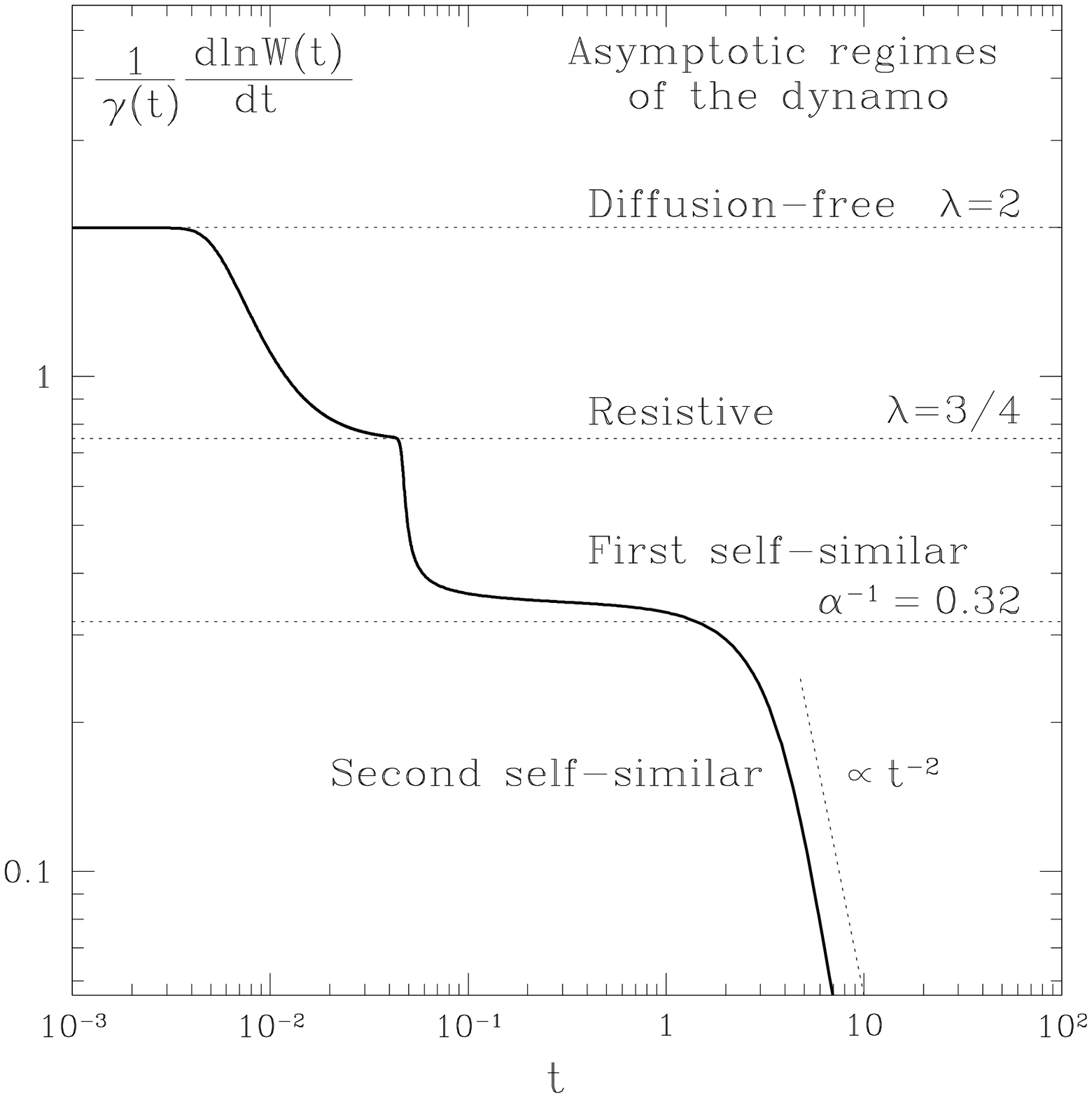,width=3in}
\epsfig{file=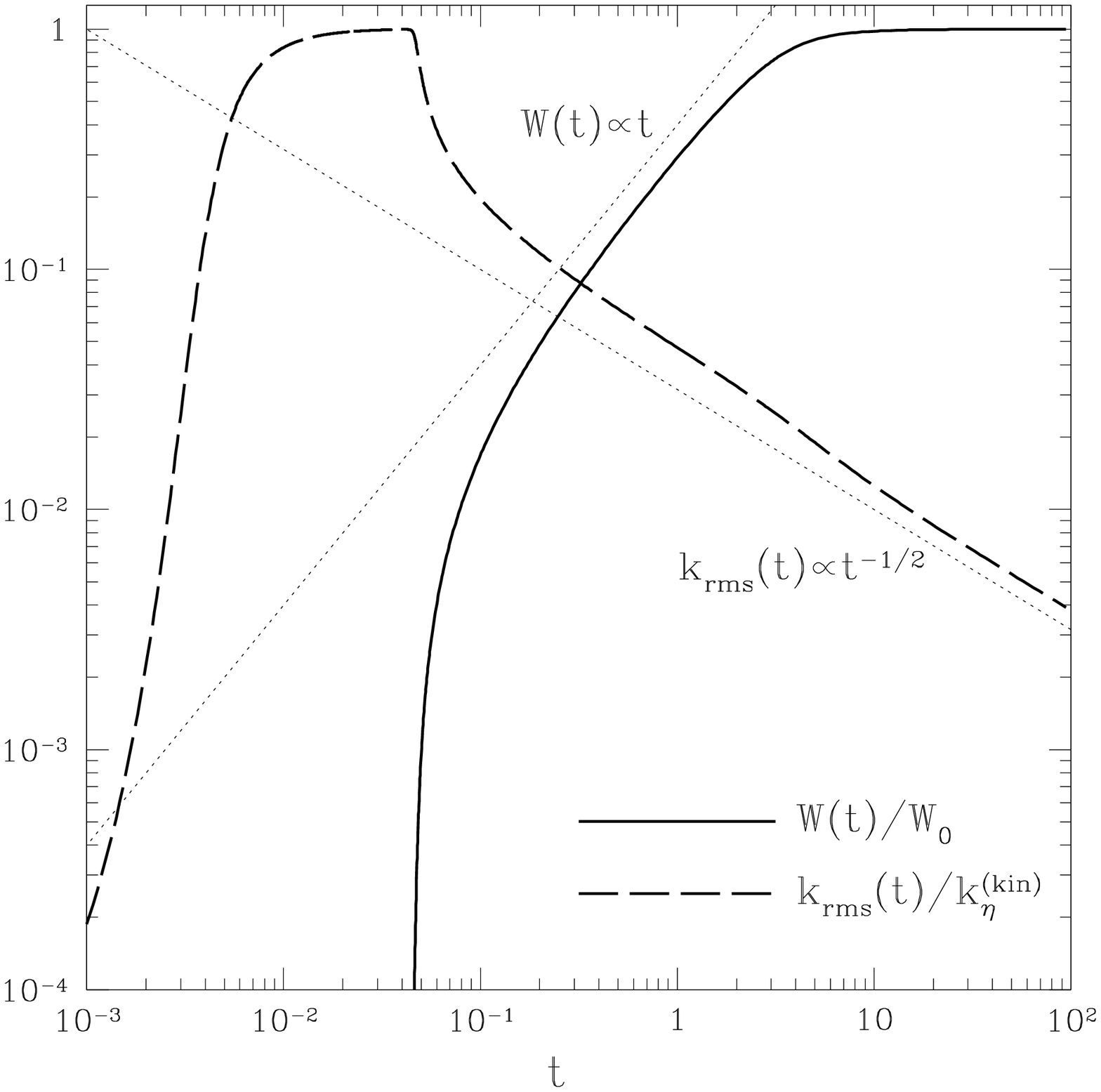,width=3in}}
\caption{\label{fig_ssim} Evolution of $W(t)$ and $\krms(t)$ 
in a numerical solution of equation~\eref{SSF_eq_nlin}. 
In order to highlight scalings,  
an extreme asymptotic regime is considered in which $\Re\sim10^7$ 
and $\Pr=10^{12}$. 
{\em Left panel:} time evolution of the quantity 
$\lambda(t)=\l[\gamma(t)\r]^{-1}{\rmd\over\rmd t}\ln W(t)$, 
which clearly exhibits all four asymptotic regimes of the dynamo: 
kinematic diffusion-free ($\lambda=2$, \ssecref{ssec_diff_free}), 
kinematic resistive ($\lambda=3/4$, \ssecref{ssec_resistive}), 
nonlinear growth/first self-similar 
($\lambda=\alpha^{-1}$, \ssecsref{ssec_nlin_growth}{ssec_first}) 
and approach to saturation/second self-similar  
($\lambda\propto t^{-2}$, \ssecsref{ssec_approach}{ssec_second}). 
{\em Right panel:} corresponding time histories 
of~$W(t)$ and~$\krms(t)$ normalized by~$W_0$ and~$\kreskin$, 
respectively. Most of the exponential growth 
of~$W(t)$ during the kinematic stage is not shown 
in order to resolve the nonlinear stages. 
The steady state (\ssecref{ssec_steady_state}), where  
the Fokker--Planck model~\eref{SSF_eq_nlin} is not strictly valid,
is not displayed either.} 
\end{figure}

The zero-flux boundary condition for the self-similar 
solution~\eref{M_ssim} has the form~\eref{zero_flux} 
with $\xiIR=\kIR/\kres(t)$. 
Substituting~\eref{Phi_sln_ssim} into~\eref{zero_flux}, 
we get the following equation for~$\alpha$: 
\bea
\l[\nu(\alpha)+\sigma(\alpha)-4\r] K_{\nu(\alpha)}(\xiIR) 
-\xiIR K_{\nu(\alpha)+1}(\xiIR) = 0.
\eea  
As explained above, this equation must be considered in 
the limit~$\xiIR\ll1$ and the resulting $\alpha$ must remain 
finite as~$\xiIR\to0$. Again there are no solutions with 
real~$\nu(\alpha)$. For imaginary~$\nu(\alpha)=\rmi\imnu(\alpha)$, 
the requirement that the spectrum should have no nodes to the right 
of~$\xiIR\ll1$ again implies~$\imnu(\alpha)\ll1$, and we~get 
\bea
\label{zero_M_ssim}
\fl K_{\rmi\imnu(\alpha)}(\xiIR)\simeq 
-{1\over\imnu(\alpha)}\,\sin\l[\imnu(\alpha)\ln(\xiIR/2)\r] = 0,\qquad 
\imnu(\alpha)\simeq 5^{1/2}6^{1/4}\({1\over\alpha_+}-{1\over\alpha}\)^{1/2}. 
\eea
The solution is (see \figref{fig_ssim})
\bea
\label{alpha_sln_first}
\alpha^{-1} \simeq \alpha^{-1}_+ - {\pi^2\over5\sqrt{6}\l[\ln(\xiIR/2)\r]^2} 
\simeq 0.32 - \Or\({1\over\ln^2(t/\tres)}\).
\eea
Since $t\ll\tres$, the spectral profile is well described 
by~\eref{Phi_sln_ssim} with $\nu(\alpha)=0$ and $\sigma(\alpha)\simeq1.9$. 

\begin{figure}[t]
\centerline{\epsfig{file=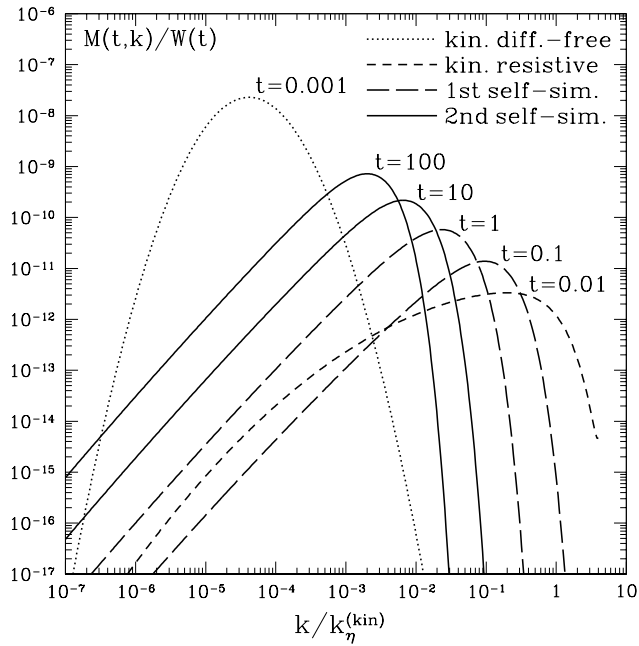,width=3.5in}}
\caption{\label{fig_Mk} Evolution of $M(t,k)$ 
in a numerical solution of equation~\eref{SSF_eq_nlin}. 
This is the same run as in \figref{fig_ssim}. 
The spectra are normalized by~$W(t)$.} 
\end{figure}

Thus, in the first self-similar stage, 
the scale-by-scale suppression of the inertial-range eddies 
leads to gradual renormalization of the 
resistive cutoff~$\kres(t)$, which is moving leftwards as~$(\eta t)^{-1/2}$ 
in agreement with the estimate in \ssecref{ssec_nlin_growth} 
(see figures~\ref{fig_ssim},~\ref{fig_Mk}). 
The turbulent power arriving 
into the smallest unsuppressed eddies is (partially) 
transfered into the small-scale magnetic fields.
The fraction~$1/2\alpha$ ($\sim16\%$) of it 
contributes to the linear-in-time growth of the total magnetic 
energy, while the rest is dissipated resistively. 
Since this stage of the dynamo lasts up to 
$\teq\sim\gKA^{-1}\Re^{1/2}$, we recover the physical estimates 
made in \ssecref{ssec_nlin_growth}: 
$W(\teq)\sim\Re^{1/2}\Wd\sim\Weq$ and 
$\kres(\teq)\sim\Re^{-1/4}\kreskin\sim\(\Re\,\Pr\)^{1/2}\kf$.\\

\noindent
{\em Remark on the elongation of the folds.} In \secref{ssec_nlin_growth}, 
we assumed the growth of the effective stretching scale~$\ls(t)$ 
during the nonlinear-growth stage to be accompanied by simultaneous 
elongation of the folds, so that the parallel scale of the magnetic field 
(characteristic length of the folds) would always 
be~$\lpar(t)\sim\ls(t)$. This assumption can now be supported 
in the following way. Consider a straight segment of a field line 
of length~$\lpar$. 
In the absence of resistivity, volume- and flux-conservation constraints 
in an incompressible flow imply $\lpar/B\sim\const$, so 
$\rmd\log\lpar/\rmd t=(1/2)\,\rmd\log W/\rmd t=\gamma(t)$. 
With resistivity included, the growth of the magnetic field is offset 
by resistive diffusion. As we showed above, 
this lowers the net amplification rate of the magnetic energy 
to $\alpha^{-1}\gamma(t)$ in the first self-similar regime. 
However, the length of the field line continues to grow at 
the rate~$\gamma(t)$, so 
$\rmd\log\lpar/\rmd t=\alpha\,\rmd\log W/\rmd t$. 
Using the nonlinear balance~\eref{nlin_balance} and 
Kolmogorov scaling for velocity gives 
$\rmd\log\lpar/\rmd\log\ls = 2\alpha/3\simeq2$. 
By itself, this would indicate that $\lpar$ grows even 
faster than~$\ls$. However, a straight section of a field 
line remains straight during stretching only as long as it is 
still short compared to the stretching-eddy size. 
Once $\lpar$ grows to be comparable to~$\ls$, it is limited 
to~$\lpar\sim\ls$ by the curvature in the stretching eddies. 
The conclusion from this argument is that 
the length of the folds can quickly adjust to the changing 
stretching scale. Note that this conclusion does not depend 
on the precise value of~$\alpha$: it is sufficient 
that~$\alpha^{-1}\le2/3$.

\subsection{The second self-similar regime: approach to saturation}
\label{ssec_second}

Thus, after time~$\teq\sim\gKA^{-1}\Re^{1/2}$, the energy of the 
small-scale magnetic fields is very close to the equipartition 
energy~$\Weq$. We observe that 
in this regime it is still possible to have a self-similar 
solution in the form~\eref{M_ssim}: the energy is now 
(nearly) constant, so~$\mu=0$ in~\eref{W_mu}. 
Accordingly, substituting~\eref{gamma_ssim} 
into~\eref{gamma_model}, we find 
\bea
\label{W_second}
W(t) \simeq \Weq \l[1-{\alpha^2\over2}\({\teq\over t}\)^2\r],\qquad 
t \gg \teq=\gKA^{-1}{\Weq\over\Wd} \sim\gKA^{-1}\Re^{1/2}.
\eea

We now set~$\mu=0$ in~\eref{Phi_eq_mu} and proceed to solve 
the resulting eigenvalue problem in exactly the same way 
as we did in \ssecref{ssec_first}. All the same formulae and  
considerations apply, except that the expressions for~$\alpha_\pm$ 
in~\eref{nu_alpha} and for~$\imnu(\alpha)$ in~\eref{zero_M_ssim} 
are now as follows: 
\bea
{1\over\alpha_\pm} = 2\(\pm2\sqrt{2/5}-1\),\qquad
\imnu(\alpha)\simeq 5^{3/4}2^{-1/4}\({1\over\alpha_+}-{1\over\alpha}\)^{1/2}. 
\eea
The new eigenvalue is 
\bea
\label{alpha_sln_second}
\alpha^{-1} \simeq \alpha^{-1}_+ - {\pi^2\over5\sqrt{5/2}\l[\ln(\xiIR/2)\r]^2} 
\simeq 0.53 - \Or\({1\over\ln^2(t/\tres)}\).
\eea
For~$t\ll\tres$, the spectrum is given by~\eref{Phi_sln_ssim} 
with~$\nu(\alpha)=0$ and $\sigma(\alpha)\simeq2.2$.

Thus, in the second self-similar stage, the magnetic 
energy is nearly constant, 
but the resistive cutoff $\kres(t)$ continues to 
decrease as~$(\eta t)^{-1/2}$, so the magnetic energy is 
redistributed from small to larger scales 
(see figures~\ref{fig_ssim},~\ref{fig_Mk}). 
This is a very slow process, but it can, in principle, go on 
for a very long time: until self-similarity is broken 
at $t\sim\tres=(10\eta\kIR^2)^{-1}$.  
Note that, according to~\eref{W_second}, the magnetic energy 
changes sufficiently slowly to make the left-hand side of 
the energy equation~\eref{W_eq_exact} subdominant in comparison 
with the field-amplification and resistive-dissipation terms 
in the right-hand side. This is consistent with the physical 
argument of \ssecref{ssec_approach}.

\subsection{Steady states at the equipartition energy and below}
\label{ssec_steady_state}

The self-similar solutions obtained above break down after a finite 
time~$\sim\tres$ due to the presence of the finite IR cutoff~$\kIR$. 
Once the resistive cutoff~$\kres$ becomes comparable to~$\kIR$, 
the system arrives at the final steady state. 
The validity of this state is questionable since 
our model was derived under the assumption that the magnetic fields 
were at smaller scales than the velocity field, i.e.~$\kres\gg\kd$. 
Furthermore, the specifications of the steady-state solution 
that the model yields clearly will have an order-one dependence 
on the value of the IR cutoff~$\kIR$, on the type 
of the boundary condition imposed, and on the form of 
the expression~\eref{gamma_model}. 
However, we proceed to discuss this solution 
because it highlights several important issues that require 
further study. 

In the steady state, $\gamma(t)\to\gamma(\infty)=\const$,  
so the corresponding solution of equation~\eref{SSF_eq_nlin} 
is simply the $\lambda=0$ eigenmode of the kinematic 
equation~\eref{SSF_eq} with $\gKA$ replaced by $\gamma(\infty)$: 
using equations~\eref{M_ev} and~\eref{Phi_sln}, we get
\bea
\label{M_st}
\fl M(k) = \const\,\xi^{3/2} K_{i\sqrt{15}/2}(\xi),\qquad 
\xi = k/\kres(\infty),\qquad 
\kres(\infty) = \l[\gamma(\infty)/10\eta\r]^{1/2}.
\eea
The value of~$\gamma(\infty)$ 
must be determined from the boundary condition at~$\kIR$: 
\bea
\label{bndry_condn_st}
\xiIR K'_{i\sqrt{15}/2}(\xiIR) 
- {5\over2}\, K_{i\sqrt{15}/2}(\xiIR) = 0,\qquad
\xiIR=\kIR/\kres(\infty).
\eea
This is an equation for~$\xiIR$, whose value in turn 
determines~$\gamma(\infty)=10\eta\kIR^2/\xiIR^2$. 
We have to pick the largest root of~\eref{bndry_condn_st} 
to ensure that the spectrum~\eref{M_st} has no nodes. 
This gives~$\xiIR\simeq0.59$, which is order one, as expected.  
Substituting~$\gamma(\infty)$ into~\eref{gamma_model}, we get 
the magnetic energy~$W(\infty)$ in the steady state: 
\bea
\label{W_curve}
\fl {W(\infty)\over\Weq} = \(1 + {\Wd\over\Weq}\)
\l\{\l[1+\(\(1+{\Weq\over\Wd}\)^2-1\){\gamma^2(\infty)\over\gKA^2}\r]^{-1/2} 
- \(1+{\Weq\over\Wd}\)^{-1}\r\},
\eea
which, using $\Weq/\Wd\sim\Re^{1/2}\gg1$ and $\gamma(\infty)/\gKA = 
10\eta\kIR^2/\gKA\xiIR^2\sim
\bigl(\kIR/\kreskin\bigr)^2\sim\(\kIR/\kd\)^2\Pr^{-1}$, 
can be rewritten as follows
\bea
\label{W_curve_Re}
{W(\infty)\over\Weq} \sim
\l[1+\({\kIR\over\kd}\)^4{\Re\over\Pr^2}\r]^{-1/2} - \Re^{-1/2}.
\eea

What is the correct choice of~$\kIR$? 
Of course, as long as the characteristic scale of the field 
is much smaller than that of all fluid motions, stretching 
or Alfv\'enic, i.e., $\kres\gg\kd$, the precise value of~$\kIR$ 
does not affect the self-similar asymptotics of 
\ssecsref{ssec_first}{ssec_second} in the leading order. 
What $\kIR$ represents is the presence in the problem of a spatial scale 
which leads to the breakdown of self-similarity in the long run 
(after time~$\tres$). As we explained in \ssecref{ssec_MHD}, 
once the resistive and the viscous scales become comparable, 
the evolution of the small-scale fields cannot be separated 
from that of the inertial-range Alfv\'enic turbulence. 
Clearly, beyond this point, the evolution of the spectrum is 
not described by the self-similar solutions obtained 
in \ssecsref{ssec_first}{ssec_second}. If we set~$\kIR\sim\kd$, 
the end of the self-similar regime and the stabilization 
of~$\kres$ at a value~$\sim\kd$ are reflected 
by the steady-state solution of our Fokker--Planck model. 
Note that, with this choice of~$\kIR$, equation~\eref{W_curve_Re} 
correctly reproduces both $W(\infty)\sim\Weq$ for~$\Pr\gg\Re^{1/2}$ 
and the physical estimate~\eref{W_subeq} 
of the subequipartition saturation level for~$\Pr\ll\Re^{1/2}$. 
We stress that, since the validity of the Fokker--Planck model 
itself breaks down together with the self-similarity, 
the specific functional form of its steady-state solution 
cannot be expected to have predictive value. 
The ensuing regime is controlled by the interaction between 
the folded fields and the Alfv\'en waves and is left 
for future study. 

\section{Discussion} 
\label{sec_conclusions}

We have presented a physical scenario of the evolution of the 
nonlinear large-$\Pr$ dynamo according to which a magnetic-energy 
spectrum concentrated at velocity scales does emerge, 
but the time scale for it is resistive, not dynamical: 
the resistive scale approaches the viscous scale 
after~$\tres\sim\(\eta\kd^2\)^{-1}$. 
In the interstellar medium, 
the Spitzer~\cite{Spitzer} value of the magnetic diffusivity 
is~$\eta\sim10^{7}~{\rm cm}^2/{\rm s}$, 
$\kd\sim10^{-16}~{\rm cm}^{-1}$~\cite{SBK_review}, so 
$\tres\sim 10^{17}$~years, 
which exceeds the age of the Universe by 7~orders 
of magnitude! The conclusion is that this mechanism cannot be invoked  
as a workable feature of the galactic dynamo --- at least not 
if the dissipation of the magnetic field is controlled by the 
Spitzer magnetic diffusivity. 



The statistically steady solutions 
discussed in \ssecref{ssec_MHD} represent 
the possible long-time asymptotic states 
of the fully developed isotropic large-$\Pr$ MHD turbulence. 
Since the approach to these states is controlled by 
the resistive time scale associated with the velocity 
spatial scales, they have practically no relevance for astrophysical 
plasmas, which have large~$\Re$ and huge~$\Pr$. 
Studying these states may, therefore, appear to be 
a purely academic exercise in fundamental turbulence theory. 
While the importance of such pursuits ought never to be 
underestimated, we should like to point out another, more 
practical angle from which our results could be viewed. 
The enormous scale separations so often encountered 
in astrophysical objects are not accessible in the turbulence 
simulations that constitute the state of the art on Earth, 
so one usually has to be satisfied with only very modest $\Re$ and~$\Pr$. 
Most of the numerical studies of the small-scale 
dynamo have so far adopted the strategy of resolving a reasonable 
hydrodynamic inertial range while only allowing for 
Prandtl numbers of at most order~10~\cite{Meneguzzi_Frisch_Pouquet,Kida_Yanase_Mizushima,Brandenburg,Chou,MCM_dynamo,Brummell_Cattaneo_Tobias}. 
In this context, our steady-state solution and formula~\eref{W_curve} 
point to an important possibility. 
When the scale separations in the problem are not very large, 
the system may converge to a steady state with a subequipartition 
magnetic energy $W(\infty)<\Weq$. Subequipartion saturated 
states have indeed been reported in the numerical experiments 
cited above. 
Since the condition for the true asymptotic regime 
is $\Pr\gg\Re^{1/2}$ (\ssecref{ssec_MHD}), 
numerical experiments with very high resolution are required 
to adequately study the large-$\Pr$ MHD turbulence. 

In conclusion, we list some of the unresolved issues that 
require further study. 

\begin{itemize}

\item The detailed mechanism of the flow suppression 
by the small-scale magnetic fields is a long-standing physical 
problem~\cite{Cattaneo_Hughes_Kim,Zienicke_Politano_Pouquet,Kim,Nazarenko_Falkovich_Galtier}. 
In our model, we have conjectured the suppression of the 
shearing flows but allowed for oscillatory Alfv\'enic 
motions, that do not stretch the field (\ssecref{ssec_nlin_growth}). 
Strictly speaking, 
only the parallel component of the shear tensor~$\bnabla\vu$ 
must be suppressed in order to block the growth of the magnetic 
energy. The perpendicular components, if unaffected, could perhaps 
mix the field (without amplifying it) 
via a quasi-two-dimensional field-line-interchange 
mechanism and thus prevent any increase in 
the field's characteristic scale. In fact, 
this seems to be the physics behind the recent numerical results on 
the large-$\Pr$ MHD turbulence {\em with a fixed uniform mean 
field}~\cite{Cho_Lazarian_Vishniac_new_regime}. 
However, it is unclear to what extent such motions can remain
quasi-two-dimensional and avoid twisting up the magnetic field,
which would then resist further mixing --- 
particularly, in our case of a folded field-line structure 
without an externally imposed mean field. 

\item Our physical picture of the final approach to saturation 
(\ssecref{ssec_approach}) 
is not based on a solid phenomenological 
argument such as that for the nonlinear-growth stage, 
and remains incompletely understood. Indeed, it remains to be 
proven definitively that the selective decay will continue after the 
small-scale magnetic energy reaches the energy of the 
outer-scale eddies. Due to resolution constraints discussed 
above, no numerical evidence of this regime is available as yet.

\item Two possibilities 
for the long-time behaviour of the isotropic MHD turbulence 
were identified in \ssecref{ssec_MHD}: saturation 
in the generic Alfv\'enic MHD turbulent state and saturation 
in a state where a significant amount of the magnetic energy 
remains tied up in the viscous-scale fields. Which one is realized 
depends on the way the small-scale folded fields interact 
with the inertial-range Alfv\'en-wave turbulence. 
We should like to observe that there is no numerical evidence 
available at present that would confirm that, the isotropic 
forced MHD turbulence {\em without externally imposed mean field} 
--- at any Prandtl number --- 
attains the Alfv\'enic state of scale-by-scale equipartition envisioned 
in Iroshnikov--Kraichnan~\cite{Iroshnikov,Kraichnan_IK} or 
Goldreich--Sridhar~\cite{Goldreich_Sridhar_strong} phenomenologies. 
In fact, medium-resolution numerical investigations~\cite{MCM_dynamo} 
rather seem to support the final states with small-scale 
energy concentration even for~$\Pr=1$. 
This does not mean that the Alfv\'enic picture is incorrect {\em per se}.
However, all existing phenomenologies of the Alfv\'enic turbulence 
depend on Kraichnan's~\cite{Kraichnan_IK} assumption 
that the most energetic magnetic fields are those at the outer scale. 
This is automatically satisfied if a finite mean field is imposed 
externally~\cite{Maron_Goldreich,Cho_Lazarian_Vishniac_GS,Cho_Lazarian_Vishniac_new_regime}. 
However, it remains to be seen if such a distribution of energy 
is set up self-consistently when the turbulence is isotropic. 
The alternative identified here involves Alfv\'enic 
motions superimposed on small-scale folded fields. 

\item An extension of the MHD description itself that may 
change the properties of the small-scale magnetic turbulence 
is the incorporation of the Braginskii~\cite{Braginskii} 
tensor viscosity~\cite{Montgomery,MCM_dynamo,Malyshkin_Kulsrud}. 
Even at magnetic energies small enough 
for the kinematic approximation to hold, the plasma is already 
well magnetized and the anisotropy of the viscous stress tensor 
leads to suppression of the velocity diffusion perpendicular 
to the local magnetic field~\cite{Malyshkin_Kulsrud}. 
This anisotropy is all the more important in view of the 
locally anisotropic (folding) structure of the magnetic 
field itself~\cite{SCMM_folding,Malyshkin_Kulsrud}. 

\item Another important extension is to allow compressible 
motions. The interstellar turbulence is sonic at the outer 
(supernova) scale but becomes subsonic and predominantly vortical 
in the inertial range (cf.~\cite{Balsara_Pouquet,Lithwick_Goldreich_compr}). 
This is the commonly accepted justification for the use of incompressible 
MHD in the models of galactic dynamo. Of course, a certain 
amount of compressive motion is always present in the interstellar 
medium (as well as in other astrophysical plasmas), 
and it has been suggested in the literature that the compressibility 
of the motions and interactions between Alfv\'en waves 
and density inhomogeneities can be important in the description 
of turbulence (see, e.g.,~\cite{Tsiklauri_Nakariakov} and many 
recent references cited therein). The treatment of the 
kinematic-dynamo problem for a model compressible case 
can be found in~\cite{SBK_review,SCMM_folding}. However, 
it remains to be understood whether and how weak compressibility 
affects the basic properties of the nonlinear dynamo. 

\end{itemize}

It is a very old observation that any extension of the domain 
of one's knowledge lengthens the perimeter 
along which this domain borders on the unknown. The perceptive 
reader must have realized that this work raises more questions 
than it gives definitive answers. We shall address these questions 
in our future investigations. 

\ack

The authors wish to thank B.~Chandran for several stimulating discussions. 
This work was supported by the UKAEA Agreement No.~QS06992, 
the EPSRC Grant No.~GR/R55344/01, and by the US-DOE Contract
No.~DE--AC02--76CHO3073. 

\appendix

\section{Alfv\'en waves and folded fields}
\label{ap_waves}

Defining $H^{ij}=B^i B^j$, we can write
the MHD equations in the following tensor form: 
\bea
\label{ui_eq}
\d_t u^i + u^k\d_k u^i = - \d_i p + \d_k H^{ik} + \nu\Delta u^i + f^i,\qquad
\d_i u^i = 0,\\
\label{Hij_eq}
\d_t H^{ij} +  u^k\d_k H^{ij} = H^{kj}\d_k u^i + H^{ik}\d_k u^j 
+ \eta R^{ij},
\eea
where $\d_k=\d/\d x^k$, 
and $R^{ij}=B^i\Delta B^j + B^j\Delta B^i$. 

\begin{figure}[t]
\centerline{\epsfig{file=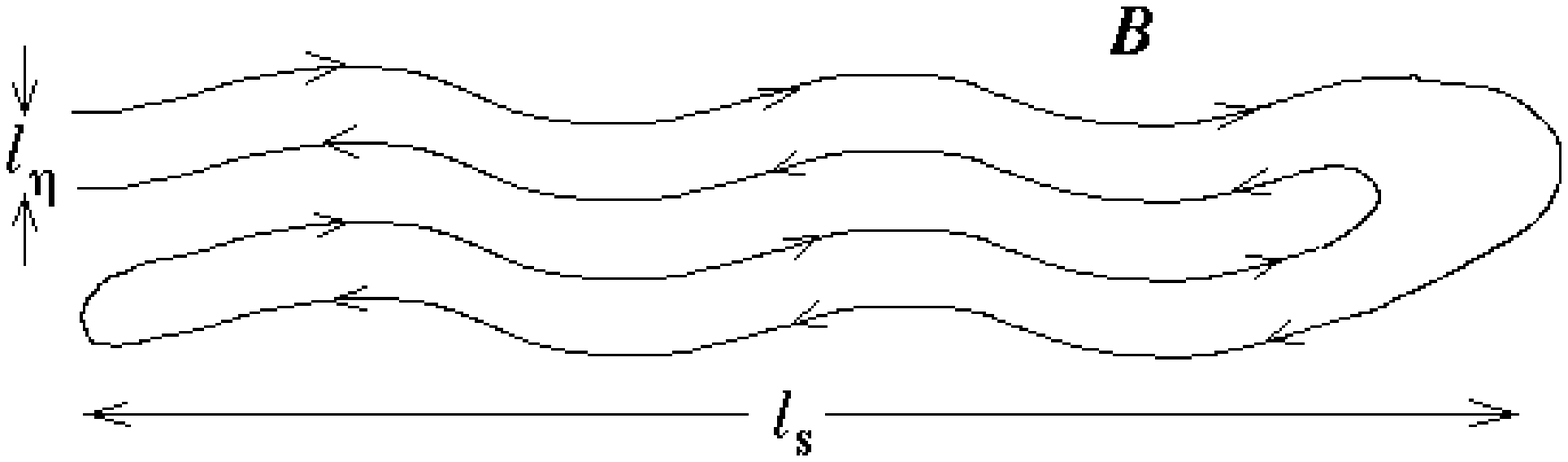,width=4in}}
\caption{\label{fig_waves} Alfv\'en waves propagate along 
folded magnetic fields.}
\end{figure}

As we explained in \ssecref{sec_phenom}, the stretching and 
folding of the magnetic field by the turbulent eddies of scale~$\ls$ 
lead to the field becoming organized in folds of length~$\lpar\sim\ls$ 
with direction reversals within each fold at 
the scale~$\lperp\sim\kres^{-1}$~\cite{SCMM_folding,SMCM_structure}. 
Consider now some typical instance during the nonlinear stage 
of the dynamo when $\kd^{-1}\ll\ls<\kf^{-1}$ (all stretching motions 
up to the scale~$\ls$ are suppressed). 
There is a clear scale separation in the problem: 
writing the magnetic-field tensor in the form 
$H^{ij}=b^i b^j B^2$, where $b^i=B^i/B$, we see that $B^2$ 
varies at scales~$\lperp\sim\kres^{-1}\ll\kd^{-1}$  
while the tensor~$b^i b^j$ varies at scales~$\lpar\sim\ls\gg\kd^{-1}$ 
($b^i$ flips its sign at scale~$\lperp$, but this does not affect the tensor 
product~$b^i b^j$). We can, therefore, average all fields over 
the subviscous scales: 
$\la u^i\ra_{\rm ss}=u^i$, 
$\la H^{ij}\ra_{\rm ss}=b^i b^j\la B^2\ra_{\rm ss}$. 
In the suppressed part of the inertial 
range, i.e. at scales~$\ell$ such that $\kd^{-1}\ll\ell\ll\ls$,  
the tensor $\la H^{ij}\ra_{\rm ss}=b^i b^j\la B^2\ra_{\rm ss}$ 
can be considered constant. Perturbing around this value and~$u^i=0$, 
we find that equations \eref{ui_eq}-\eref{Hij_eq} (without 
the dissipation terms) admit wave-like solutions with the dispersion 
relation 
\bea
\label{disp_rln}
\omega=\pm\kpar\vA,\qquad 
\kpar=\sqrt{k_i k_j b^i b^j},\qquad 
\vA=\sqrt{\la B^2\ra_{\rm ss}}.
\eea
These are Alfv\'en-like waves propagating along the folds 
of the direction-reversing magnetic fields, which are straight 
at scales below~$\ls$ (\figref{fig_waves}). These waves are a modification 
of the previously introduced {\it magnetoelastic 
waves}~\cite{Gruzinov_Diamond,Chandran_thesis} with account taken 
of the folding structure of the small-scale magnetic fields. 

In~\eref{disp_rln}, we must have 
$\ls^{-1}\ll\kpar\ll\kd$, so the frequency~$\omega$ is 
larger than the characteristic time scale of the growth of 
the small-scale magnetic energy: 
$\omega =\kpar\vA\gg\vA/\ls\sim W^{1/2}/\ls\sim u_{\ls}/\ls\sim\gamma$. 
Since the resistive-dissipation rate of the small-scale 
fields is~$\eta\krms^2\sim\gamma$ by virtue of~\eref{W_eq_exact}, 
the Alfv\'en waves are mostly dissipated {\em viscously} 
via the MHD turbulent cascade. 
The velocity spectrum of this Alfv\'enic turbulence is subject to 
considerations similar to those arising in 
the Iroshnikov--Kraichnan~\cite{Iroshnikov,Kraichnan_IK} or 
the Goldreich--Sridhar~\cite{Goldreich_Sridhar_strong} 
phenomenologies. Further investigations of this and related issues 
will be published elsewhere.

\end{document}